\documentclass{jetpl}
\twocolumn
\usepackage{amsmath,amsfonts,epsfig,graphics,amssymb}

\usepackage{hyperref}
\hypersetup{colorlinks=true,citecolor=blue}

\lat
\title{Corona effect in AA collisions at LHC}
\rtitle{Corona effect in AA collisions at LHC \ldots}
\sodtitle{Corona effect in AA collisions at LHC}
\author{V.\,S.\,Pantuev \/\thanks{e-mail: pantuev@inr.ru}}

\rauthor{V.\,S.\,Pantuev}
\sodauthor{Pantuev}
\address{Institute for Nuclear Research RAS, 117312 Moscow, Russia}

\dates{}

\begin{document}
\abstract{Following our earlier finding based on RHIC data on the dominant jet production from nucleus corona region, we reconsider this effect in nucleus-nucleus collisions at LHC energies. Our hypothesis was based on experimental data, which raised  the idea of a finite formation time for the produced medium.  At RHIC energy and in low density corona region this time reaches about 2~fm/$c$.  Following this hypothesis, the nuclear modification factor $R_{AA}$ at high $p_t$ should be independent on particle momentum, and the azimuthal anisotropy of high $p_t$ particles, $v_2$,  should be finite. A separate prediction held that,  at LHC energy, the formation time in the corona region should be about 1~fm/$c$.  New data at LHC show that $R_{AA}$ is not flat and is rising with $p_t$. We add to our original hypothesis an assumption that a fast parton  traversing the produced medium loses the fixed portion of its energy. A shift of about 7~GeV from the original power law $p^{-6}$ production cross section in $pp$ 
explains well all the observed  $R_{AA}$ dependencies. The shift of about 7~GeV is also valid at RHIC energy. We also show that the observed at LHC dependence  of $v_2$ at high $p_t$ and our previous predictions agree.
}


\maketitle
\begin{figure}[thb]
\centering{\includegraphics[width=0.8\linewidth]{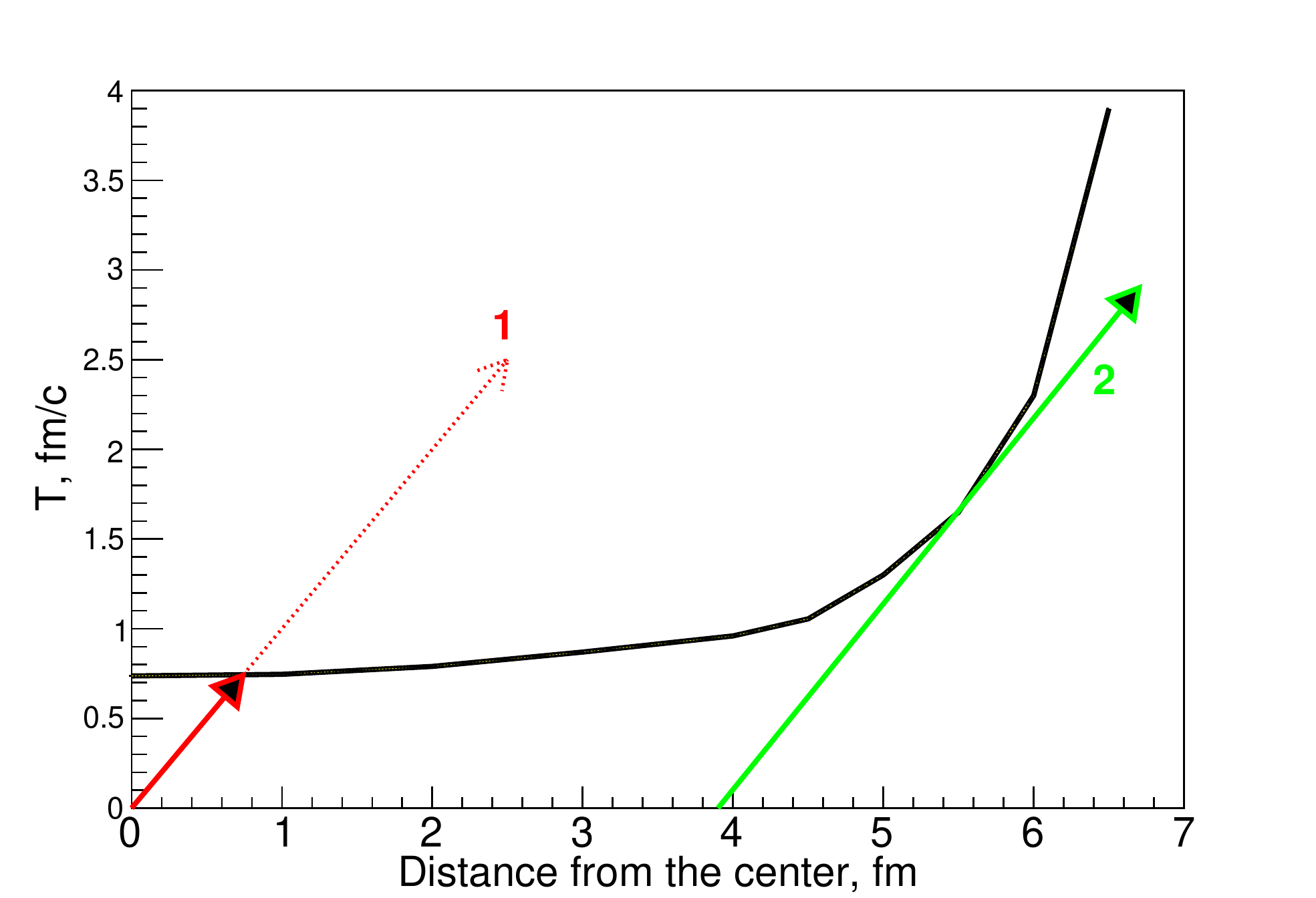}}
\caption{\label{fig:density} Fig.1. The value of formation time versus distance from the center of the colliding region, solid line, in most central AuAu collisions at $\sqrt s_{\rm NN}$=200~GeV at RHIC. The arrows demonstrate world lines for two fast partons moving with the speed of light. The first parton was produced right in the center, the second -- near the surface.}
\end{figure}

Over the last 17 years of relativistic nucleus-nucleus collisions at RHIC and LHC, a set of observables was found which confirms the formation of high energy and high density matter. Among these features are the strong jet suppression manifested in particle suppression at high transverse momentum, $p_t$, and large particle anisotropy. There is also a long list of models and theoretical assumptions to explain these effects. In our view, when one talks about  jet suppression, a significant effect of particle production from the nucleus corona region is often ignored or underestimated. In a previous publication based purely on experimental data at RHIC, a simple model was proposed~\cite{jetp_l_2007} to explain the angular dependence in the reaction plane of the nuclear modification factor $R_{AA}$. The model nicely described the centrality  and azimuthal  dependence (or factor $v_{2}$ for high $p_t$) of $R_{AA}$ at RHIC energy. In the model, there is one free parameter of about 2.3~fm/$c$ which was  interpreted as plasma formation time at the low density corona region. The physical meaning of this parameter is that fast partons have roughly this time to escape from the produced medium and, theafter, they are absorbed by the absolutely opaque central region. This value of $T_0$=2.3~fm/$c$ is not ``crazy large'' because the number of nuclear collisions, $N_{coll}$, near corona region is rather small, but it should be less than 0.8~fm/c in the core region of the produced matter~\cite{density}. Time,
necessary to form the strongly interacting colored matter, should be proportional to the mean distance between
the interaction or collision points with a color exchange.  This distance, itself, should be inversely proportional to the square root of the density of such interactions. The picture in some sense is similar to the percolation scenario~\cite{satz}.  If the density  of $N_{coll}$ in $x-y$ plane of colliding nuclei near the corona region is $\rho_{periph}$ then the formation time $T(r)$ versus density $\rho(r)$ will be:
\begin{equation}  
T(r)=T_0\cdot \sqrt{\rho_{periph}/\rho(r)},
\label{eq:T_r}
\end{equation}
where $r$ is the distance from the center.
In Fig.~\ref{fig:density} we plot the evolution of the formation time versus the distance from the center of the region for the colliding Au+Au nuclei in the 0-5\% centrality bin. For the $N_{coll}$ density distribution of colliding nucleons we used density profiles generated for our first publication~\cite{jetp_l_2007}. To demonstrate how formation time works we show in Fig.~\ref{fig:density} two extreme cases: when the fast parton is produced in the center of the colliding region, arrow 1, and near the surface at a depth of about 2 fm from the Woods-Saxon radius, arrow 2. The first parton moves with the speed of light along its world line 1 only for about 0.8~fm and then is stopped by the produced matter. The second parton will survive. The proposed model in~\cite{jetp_l_2007} works well at RHIC energy.

In ``The last call for prediction'' published prior to the start of LHC we also proposed some features which should be observed at LHC if a similar picture with formation time is valid~\cite{last_call} (see  pages 119--121 and figures   99--100 in the e-print version).   As we already mentioned, the formation time should be proportional to the mean distance between
interactions with the color exchange. It means also that only part of the nucleon-nucleon inelastic cross section will contribute to the process:  single- and double diffractive and soft process with meson exchange will not be relevant here. If a relativistic rise
of the total nucleon-nucleon, NN,  cross section comes purely from the contribution from the colored parton hard scatterings,
we can estimate the relative value of hard scatterings to the total
 nucleon-nucleon cross section. At $\sqrt(S_{NN})$=20 GeV the NN total cross section is at its
minimum of  30 millibars  - there is almost no hard scattering, but  mostly soft nucleon-nucleon interactions with meson exchange. At  200 GeV the cross section rises by 13~mb, at 5500 GeV -- by 49~mb. The formation time of the colored matter 
should be proportional to one over the square root of these numbers because the density of $N_{coll}$ is proportional to the cross section. If we get 
$T$=2.3~fm/$c$ at 200~GeV, then from the rise of the NN total cross section, we estimate T=1.2~fm/$c$ near the corona region at around 5 TeV of  LHC energy. In the center of the collision zone it will be about three times shorter. Calculations show that such a value of $T$ should give a constant $R_{AA}$=0.1 for high $p_t$ particles in the most central collisions. We have to emphasize that the value of  T around 1.2~fm/$c$ is valid within uncertainty of 5\%  in the LHC energy range of 2.7--5 TeV. It comes  from a little change of $pp$ total cross section between 85~mb and 90~mb if one interpolates the existing $pp$ data~\cite{pdg}, thus, and relative change of hard scattering contribution is on the level of 5 mb.

\begin{figure}[thb]
\centering{\includegraphics[width=0.8\linewidth]{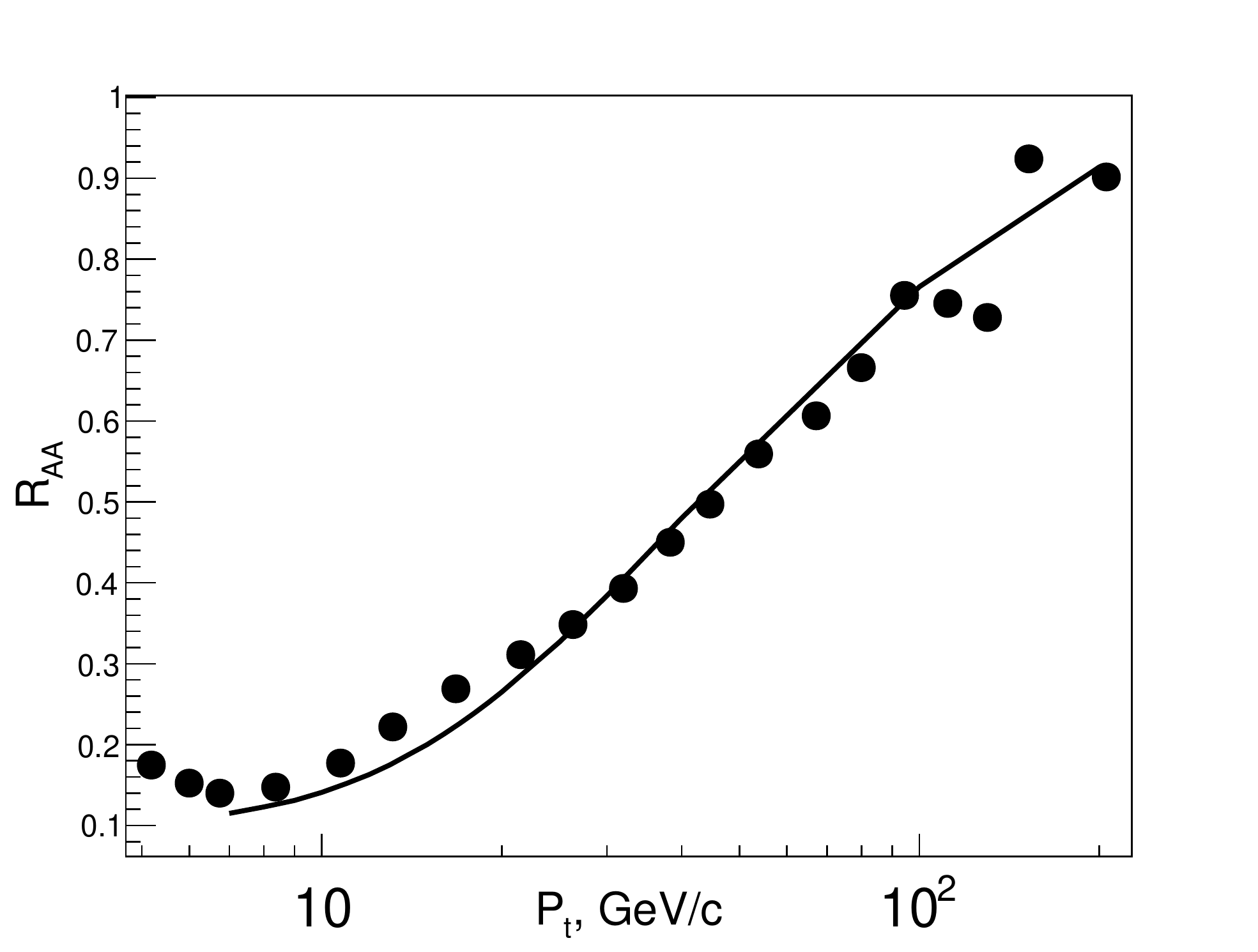}}
\caption{\label{fig:cms_0-5} Fig.2. The dependence of single particle $R_{AA}$ versus transverse momentum  $p_t$. The points are data from the CMS collaboration for the most central 0-5\% PbPb collisions at $\sqrt s_{\rm NN}$=5.02~TeV~\cite{cms}. The line is our estimation.}
\end{figure}

\begin{figure}[thb]
\centering{\includegraphics[width=0.8\linewidth]{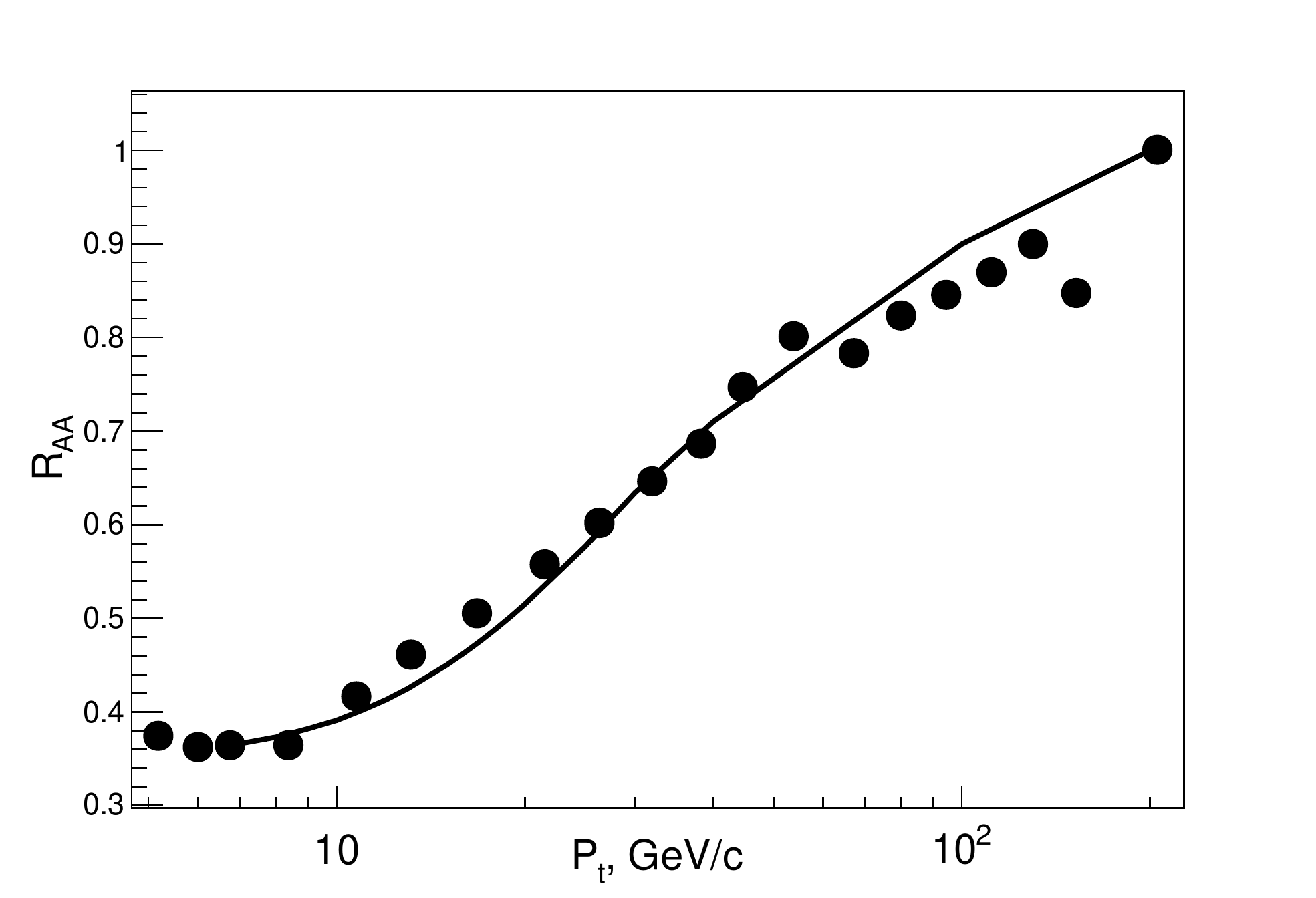}}
\caption{\label{fig:cms_30-50} Fig.3. The same as Fig.~\ref{fig:cms_0-5} but for centrality 30-50\% of PbPb collisions at $\sqrt s_{\rm NN}$=5.02~TeV~\cite{cms}.}
\end{figure}

Predictions made in~\cite{last_call} assume that the core of the produced matter is opaque, but experimental data for Pb+Pb collisions obtained by  ALICE,  CMS and ATLAS~\cite{alice_pt, cms_pt, atlas_pt} show that $R_{AA}$ is continuously rising at high $p_t$. It means that the core of the collision zone becomes more transparent for fast particles. It is natural to assume that the parton loses some portion of its energy. We found that a constant energy loss of 7~GeV describes well the data for $R_{AA}$ versus $p_t$. Particle, namely pion,  production cross section at LHC energy follows a simple power law $p^{-6}$~\cite{alice} at high $p_t$. Thus, the energy drop by 7~GeV becomes less significant with increasing  parton $p_t$. In Fig.~\ref{fig:cms_0-5} we present results for the $R_{AA}$ versus $p_t$ from CMS data~\cite{cms} and our calculations for most central collisions. There are two contributions: a constant value of 0.1 for a particle from the corona region, as was predicted in ref.~\cite{last_call}, and a new momentum dependent component when matter becomes more transparent for fast parton, which loses 7~GeV.  This provides excellent agreement with the data. In Fig.~\ref{fig:cms_30-50} we show a similar plot for mid-central collisions. In this case the  contribution to $R_{AA}$ from the corona region reaches 0.35~\cite{last_call}, but the penetrating parton  contribution is about the same. 

\begin{figure}[thb]
\centering{\includegraphics[width=0.8\linewidth]{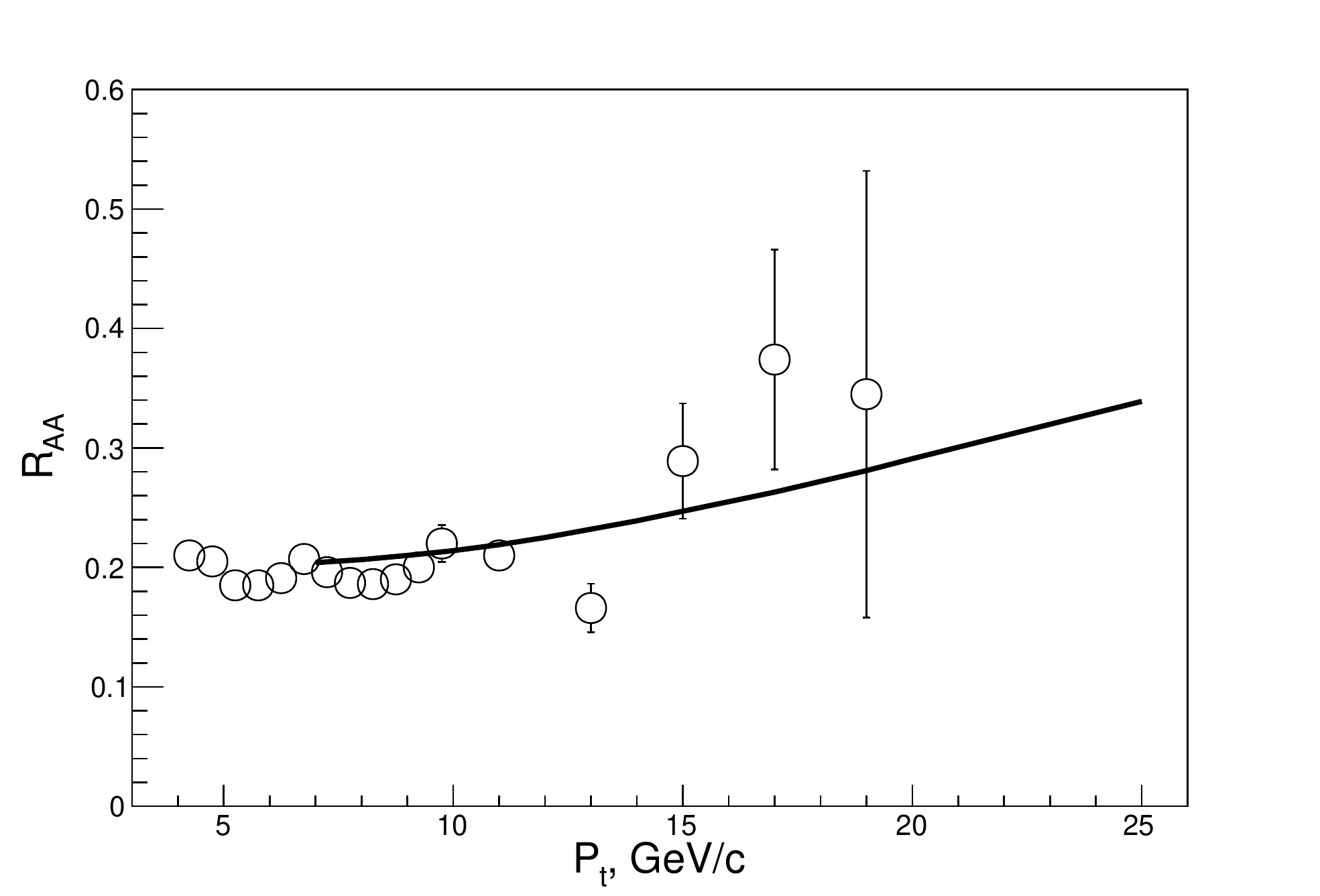}}
\caption{\label{fig:phenix_0-10} Fig.4. Re-estimation of PHENIX data for $\pi^0$  $R_{AA}$ in 0-10\% centrality bin at $\sqrt s_{\rm NN}$=200~GeV~\cite{phenix} by using the same parton energy loss of 7~GeV as at LHC energy. }
\end{figure}

Out of curiosity we checked how this 7~GeV energy loss works at RHIC and added  this component to the previous calculation with the corona region and absolutely black core, Fig.~\ref{fig:phenix_0-10}. The only difference here is that the production cross section at RHIC follows a more steep power law $p^{-8}$~\cite{phenix}. Within the error bars our line follows the experimental points. Such a large energy loss (7~GeV) at RHIC explains why the assumption about the complete black core with some corona contribution worked so well -- the loss is too big for produced particles at RHIC.

\begin{figure}[thb]
\centering{\includegraphics[width=0.8\linewidth]{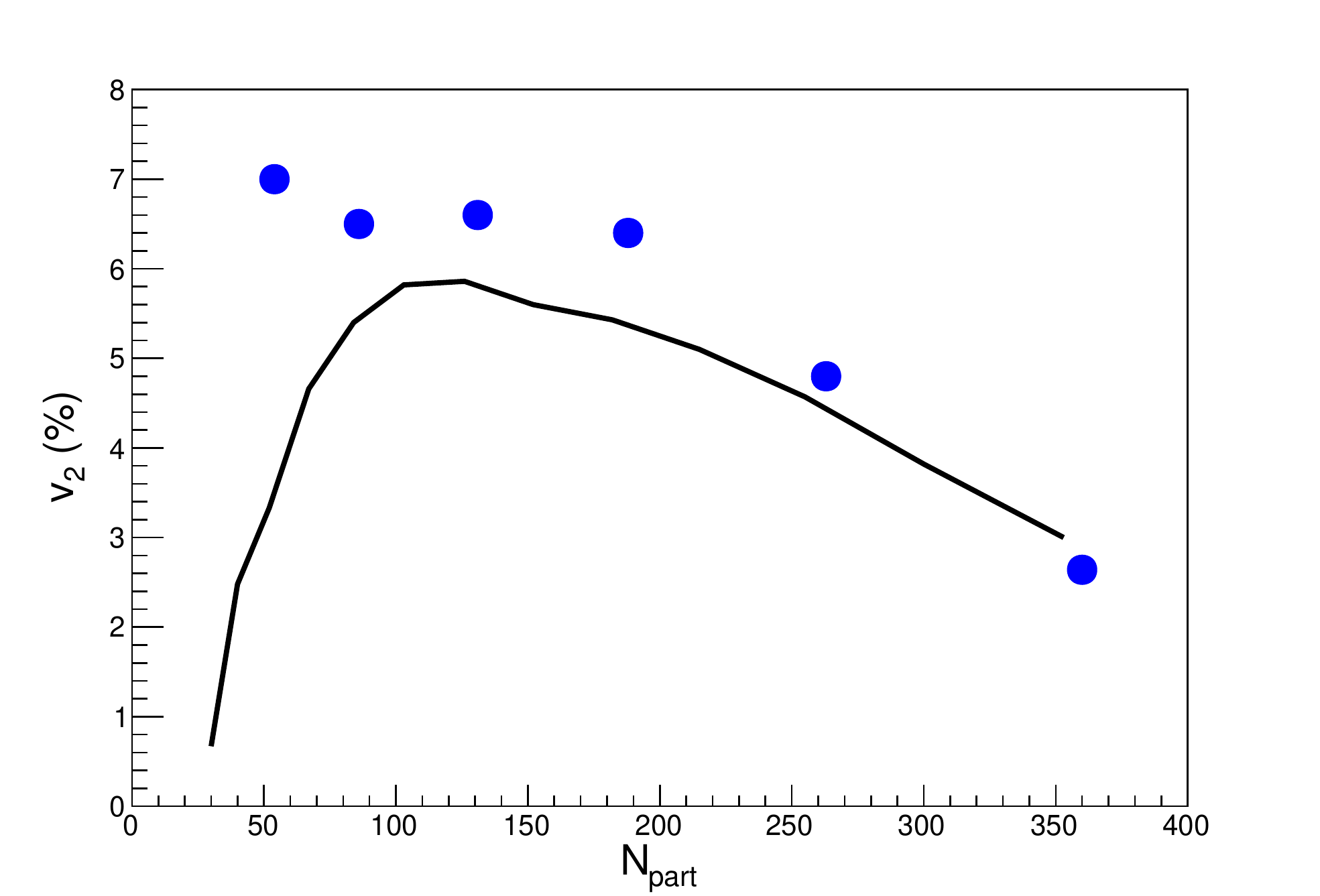}}
\caption{\label{fig:LHC_v2} Fig.5 Dependence of azimuthal asymmetry parameter $v_2$ versus number of participant nucleons, $N_{part}$. Solid line is our prediction from ref.~\cite{last_call}, points are CMS data at $p_t$=15~GeV/$c$ and $\sqrt s_{\rm NN}$=2.76~TeV~\cite{cms_v2}.}
\end{figure}

Our model worked well at RHIC for the observed large azimuthal of high  $p_t$ particles or parameter $v_2$. Nearly 10 years ago we did a prediction for
$v_2$ at LHC~\cite{last_call}. It seems that the prediction is valid. In Fig.~\ref{fig:LHC_v2} we compare our estimations with CMS results at $p_t$=15~GeV/$c$~\cite{cms_v2}. The prediction of a large $v_2$ even at LHC is confirmed, the sensitivity to the collision geometry persists up to high $p_t$. There is a deviation at small $N_{part}$ but this is due to the well known effect of distortion by the initial  geometry fluctuations (see for example, PHOBOS paper~\cite{phobos}). We also can explain the observed drop of $v_2$ with particle or jet momentum above 15~GeV/$c$. The corona effect for in- and out-of-plane particle production is diluted by penetrating partons with energy loss. For example, looking at Fig.~\ref{fig:cms_0-5} and Fig.~\ref{fig:cms_30-50}, one can see that at $p_t$=40~GeV/$c$ particles from the corona region count for about one half of the total yield at this momentum. Thus,  $v_2$ should drop to about a factor of 2. This what is qualitatively seen by the three experiments~\cite{cms_v2, ALICE_v2, ATLAS_v2}.

In conclusion, we demonstrate that in PbPb collisions at LHC the contribution from the corona region and the assumption of a finite formation time for the colored strongly interacting matter are the reasons for the observed centrality and momentum dependence of particle $R_{AA}$. At LHC energies, a fast parton escapes the interaction zone by losing about 7~GeV. Within our model this value does not depend on momentum, centrality, energy density, and, probably, on beam energy. The observed azimuthal angular asymmetry at a high transverse momentum is well described at RHIC and LHC energies.

This work was partially supported by the RFBR grant numbers 14-22-03069-ofi-m and  14-02-00570-a. We would like to thank Keith Guzik for help with the text.

\end{document}